\documentclass{article}
\usepackage{smc2026}
\usepackage{paralist}
\usepackage[figure,table]{hypcap}
\usepackage{multirow}
\usepackage{graphicx}
\usepackage{subcaption}
\usepackage{booktabs}
\usepackage{xcolor}
\usepackage{amsmath}
\usepackage{subfiles}

\usepackage{todonotes}

\newcommand{\etal}{\textit{et~al.}}
\newcommand{\ie}{\textit{i.e.}~}
\newcommand{\eg}{\textit{e.g.},~}

\newcommand{\allinone}{All-In-One}

\DeclareMathOperator*{\Fzf}{F_{0.5s}}

\DeclareMathOperator*{\Fth}{F_{3s}}

\usepackage[english]{babel}

\def\papertitle{Unsupervised Evaluation of Deep Audio Embeddings for Music Structure Analysis}
\author[1]{\mbox{\firstname{Axel}\lastname{Marmoret}\email{axel.marmoret@imt-atlantique.fr}\orcid{0000-0001-6928-7490}}}

\affil[1]{\department{BRAIN team, MEE}\institution{IMT Atlantique}\city{Brest}\country{France}\affiliationtype{University}}

\completesetup

\title{\papertitle}
\begin{document}
	\capstartfalse
	\maketitle
	\capstarttrue
		\begin{abstract}
Music Structure Analysis (MSA) aims to uncover the high-level organization of musical pieces. State-of-the-art methods are often based on supervised deep learning, but these methods are bottlenecked by the need for heavily annotated data and inherent structural ambiguities. In this paper, we propose an unsupervised evaluation of nine open-source, generic pre-trained deep audio models, on MSA. For each model, we extract barwise embeddings and segment them using three unsupervised segmentation algorithms (Foote's checkerboard kernels, spectral clustering, and Correlation Block-Matching (CBM)), focusing exclusively on boundary retrieval. Our results demonstrate that modern, generic deep embeddings generally outperform traditional spectrogram-based baselines, but not systematically. Furthermore, our unsupervised boundary estimation methodology generally yields stronger performance than recent linear probing baselines. Among the evaluated techniques, the CBM algorithm consistently emerges as the most effective downstream segmentation method. Finally, we highlight the artificial inflation of standard evaluation metrics and advocate for the systematic adoption of ``trimming'', or even ``double trimming'' annotations to establish more rigorous MSA evaluation standards.
        
	\end{abstract}

	\section{Introduction}\label{sec:introduction}
Beyond just sound, music is an intricate organization of scales and rhythms. At its core, musical composition relies on repetition, contrast, and variation to organize sound into coherent forms, establishing recognizable sections separated by distinct transitions. Music Structure Analysis (MSA) is the task of identifying this high-level organization. Specifically, it aims to partition a musical piece into meaningful, non-overlapping sections (\eg intro, verse, chorus, or bridge), and, in particular, to locate their precise temporal boundaries~\cite{nieto2020segmentationreview}. Beyond its established value in guiding applications like music summarization, cover song identification, and data-driven musicology, we believe that MSA will find renewed significance in the era of generative AI. By providing a structural framework, accurate MSA can enhance the editability, user control, and long-term coherence of AI-generated audio~\cite{morreale2025reductive, choi2025understanding}. 

Yet, despite its practical importance, MSA remains a notoriously challenging task: musical form is inherently hierarchical, and multiple structural interpretations can be simultaneously valid depending on the listener's focus, the annotation protocol, and the desired level of granularity. Early work on MSA focused on self-similarity matrices (SSMs)~\cite{nieto2020segmentationreview, foote2000automatic}, \ie square matrices indicating the pairwise similarity of all time instances in a song. From this viewpoint, sections are often understood as large, homogeneous and/or repeating regions, and boundaries as salient ruptures between subsequent sections. Comparing time instances to detect change points was the root of MSA for years, with significant research in the previous decade dedicated to refining the similarity computation between these instances~\cite{serra2014unsupervised, mcfee2014analyzing, marmoret2023barwise}. This line of work was significantly advanced by the rise of deep learning, particularly representation learning~\cite{mccallum2019unsupervised, salamon2021deep, wang2021supervised, buisson2024self}. Researchers have begun to use deep models in conjunction with standard downstream self-similarity segmentation algorithms to improve audio understanding. For instance, Salamon \etal~\cite{salamon2021deep} used deep embeddings to replace the traditional acoustic features in~\cite{mcfee2014analyzing}, improving the performance.

Parallel to these developments, supervised deep learning models~\cite{grill2015cnn, kim2023all, buisson2024using, korzeniowski2025simple} emerged as a dominant approach, explicitly learning to predict boundaries from labeled data and achieving state-of-the-art performance in MSA. However, this success comes with significant trade-offs. Supervised models rely heavily on large, meticulously annotated datasets, which are time-consuming and costly to produce. Furthermore, the inherent ambiguity of musical structure becomes a major bottleneck: supervised training biases estimates towards a single ``ground truth'' annotation protocol. This effectively discards other potentially valid, equivalent structural interpretations and limits the model's ability to generalize across different musical distributions.

Recent Self-Supervised Learning (SSL) paradigms offer a promising path to alleviate these issues. By leveraging pretext tasks (such as Masked Acoustic Modeling), 
these networks bypass the annotation bottleneck and learn directly from massive, unlabeled audio corpora. SSL dedicated to MSA (notably via MSA-specific inductive biases) has been shown to be effective in recent work~\cite{buisson2024self}. Such specialized pretraining, however, requires a tailored architecture and training pipeline dedicated to structure. This raises an intriguing open question: do broad, general-purpose audio representations naturally encode structure-related information as a byproduct of their pre-training, even without being explicitly designed for structure? There is reason to expect so, as their pre-training tasks require capturing rich timbral, temporal, and contextual characteristics. If they do, MSA could be performed without any task-specific training while directly benefiting from the rapid progress of general audio foundation models. Recent work by Toyama \etal~\cite{toyama2026foundational} began to address that question by exploring whether generic deep audio models (SSL-based but not only) could benefit MSA. However, their study evaluated learned representations using linear probing, which involves training a supervised linear head on top of the frozen embeddings to estimate structure. While computationally efficient, linear probing remains fundamentally supervised and thus sensitive to the annotation drawbacks discussed above.

Our work shares the fundamental premise of Toyama~\etal~\cite{toyama2026foundational}, but we employ strictly unsupervised downstream segmentation. In this paper, we study unsupervised boundary estimation using frozen embeddings extracted from nine open-source, generic, pre-trained deep learning audio models~\cite{huang2022masked, MERT, won2024foundation, zhu2025muq, niizumi2024m2d, quelennec2025matpac_plus, kumar2023high, pasini2025codicodec, CLAP}. Following our previous work~\cite{marmoret2023barwise}, these embeddings are computed at the bar scale. We evaluate these embeddings by computing barwise SSMs, and segment them using three standard and state-of-the-art unsupervised segmentation algorithms: Foote's checkerboard kernels~\cite{foote2000automatic}, spectral clustering (LSD)~\cite{mcfee2014analyzing}, and the Correlation Block-Matching (CBM) algorithm~\cite{marmoret2023barwise}. We focus on the boundary retrieval task only, and do not attempt to label the sections. Our goal is two-fold: on one hand, to assess to what extent modern deep audio embeddings are rich enough to provide structured information about the data, and in turn improve unsupervised MSA; on the other hand, to compare the potential of traditional unsupervised segmentation algorithms to be improved when complemented with deep representation learning.

The remainder of this paper details our methodology for extracting and segmenting barwise embeddings (Section~\ref{sec:methodology}), outlines the experimental setup and evaluation metrics (Section~\ref{sec:experiments}), discusses our comparative results (Section~\ref{sec:results}), and concludes (Section~\ref{sec:conclusion}).

\section{Methodology}
\label{sec:methodology}

\subsection{Deep Audio Models}
\label{sec:deep_models}
To evaluate the structural information captured by modern representation learning, we extract embeddings from nine distinct open-source deep audio models~\cite{huang2022masked, MERT, won2024foundation, zhu2025muq, CLAP, niizumi2024m2d, quelennec2025matpac_plus, kumar2023high, pasini2025codicodec}. These deep audio models differ significantly in their architectures and training objectives. Three of these nine models (MERT~\cite{MERT}, MusicFM~\cite{won2024foundation}, and MuQ~\cite{zhu2025muq}) are only trained on music data, and their respective authors explicitly noted their potential utility for MSA.

The majority of the evaluated architectures rely on Masked Acoustic Modeling, an SSL paradigm where the network optimizes the reconstruction of masked portions of the audio signal. Within this category, AudioMAE~\cite{huang2022masked} learns by reconstructing masked spectrogram patches. Meanwhile, MERT~\cite{MERT}, MusicFM~\cite{won2024foundation}, and MuQ~\cite{zhu2025muq} learn to reconstruct indirect representations of the audio. Specifically, MERT uses pseudo-labels from acoustic teacher models, MusicFM predicts targets via a random projection quantizer, and MuQ predicts discrete tokens generated by a single-layer Residual Vector Quantization (RVQ). Additionally, M2D~\cite{niizumi2024m2d} and MATPAC++~\cite{quelennec2025matpac_plus} employ Masked Latent Prediction, using two models: one computing latent features of masked patches, the other one estimating these latent features from visible patches. MATPAC++ further incorporates Multiple Choice Learning to model prediction ambiguity, which could turn out to be very relevant for tackling the inherent ambiguity of MSA.

A second category of evaluated models consists of Neural Audio Codecs, which are generally designed as autoencoders to compress audio as aggressively as possible while maintaining high reconstruction fidelity. The extreme compression bottleneck forces these networks to learn highly abstract and compact latent representations of the acoustic space. DAC~\cite{kumar2023high} is a prominent example of this approach, operating as a CNN-based autoencoder that utilizes RVQ to compress audio into discrete hierarchical tokens. To address the codebook collapse sometimes associated with traditional RVQ, CoDiCodec~\cite{pasini2025codicodec} utilizes Finite Scalar Quantization to project continuous representations onto a fixed integer grid. This architecture allows CoDiCodec to simultaneously extract both continuous summary embeddings and discrete acoustic tokens.

Finally, we evaluate models trained with cross-modal contrastive learning, which aligns audio and text latent spaces to bridge acoustic features with semantic concepts. CLAP~\cite{CLAP} is the only evaluated model not pretrained via SSL reconstruction objectives: its audio encoder was first trained for supervised audio classification, then aligned with text via contrastive learning. M2D~\cite{niizumi2024m2d} and MuQ~\cite{zhu2025muq}, while primarily focused on acoustic reconstruction, also incorporate a contrastive objective; both therefore yield two distinct representations: a unimodal audio space and a joint multimodal one.

\subsection{Barwise Processing}
\label{sec:barwise}

In the context of Western modern music, we adopt the bar-scale as a musically motivated temporal resolution for MSA. Musical form is fundamentally tied to this metrical hierarchy, as macroscopic structural changes and repetitions frequently align with bar boundaries. Beyond this musical intuition, the validity of downbeat-synchronized boundaries is supported empirically: our previous work~\cite{marmoret2023barwise} demonstrated that barwise alignment maintains or improves the performance of Foote's checkerboard kernels~\cite{foote2000automatic} and CBM~\cite{marmoret2023barwise} algorithms, when compared with beat-scale processing. In that sense, embeddings will be individually computed on the different bars of the signal, thus resulting in barwise embeddings.

\subsection{Downstream Segmentation Algorithm}
\label{sec:seg_algos}
After the computation of barwise deep embeddings, we apply three state-of-the-art unsupervised algorithms to perform the boundary retrieval task. All of these algorithms are based on SSMs\footnote{Authors of~\cite{mcfee2014analyzing} prefer to mention ``affinity matrices'', with a clear definition of their computation, but the main principle remains the same.}. Foote's algorithm~\cite{foote2000automatic} is a novelty-based method that identifies abrupt structural transitions by correlating a square checkerboard kernel along the main diagonal of the SSM. Conversely, the LSD approach~\cite{mcfee2014analyzing} takes a graph-theoretic perspective to capture long-term repetitions; it treats the SSM as an adjacency matrix and analyzes the eigenvectors of its normalized Laplacian to estimate boundaries via K-means change-points. Finally, the CBM algorithm~\cite{marmoret2023barwise} frames boundary retrieval as a global optimization problem solved via dynamic programming. Conceptually similar to Foote's method but optimized for detecting homogeneous regions rather than abrupt novelty, CBM identifies optimal boundaries by maximizing a score function that represents block structures around the main diagonal. 

\subsection{Barwise TF -- Baseline Representation}
\label{sec:baseline}
To situate deep embeddings against a non-deep-learning reference, we compare them to a baseline built from classical acoustic features. MSA has traditionally relied on low-level features such as Mel spectrograms, chromagrams, or MFCCs~\cite{nieto2020segmentationreview}. We adopt the \emph{Barwise TF} matrix~\cite{marmoret2023barwise}, which represents the signal as a sequence of barwise vectors, each flattening the intra-bar time-frequency content of a bar into a single vector. Following~\cite{marmoret2023barwise}, this content is computed on Log-Mel spectrograms, a representation shown in that work to perform strongly with the Foote and CBM algorithms, then reaching levels close to those of supervised deep-learning models~\cite{mccallum2019unsupervised, wang2021supervised, salamon2021deep}.

\section{Experimental Settings}
\label{sec:experiments}
\subsection{Implementation Details}

We operate entirely at the bar scale, estimating downbeats using the Beat This!~\cite{foscarin2024beat} model. To ensure compatibility, raw audio signals are segmented into bar-length chunks before computing embeddings from each model's final latent layer. Because the evaluated models operate at different native temporal resolutions, we homogenize the representations by averaging across the temporal dimension, yielding a single vector per bar. We adopt this framework for consistency, despite the contradiction between the gains reported in~\cite{toyama2026foundational} and the temporal preservation favored by~\cite{marmoret2023barwise}. Investigating this specific impact is beyond our current scope.

Using these barwise embeddings, we compute barwise SSMs via both RBF and Cosine similarity measures, then segmented using Foote's and CBM algorithms. The exception is the LSD algorithm, which inherently computes its own specialized affinity matrix to emphasize diagonal stripes. The same SSM computation is applied to the Barwise TF baseline representation introduced in Section~\ref{sec:baseline}.

Regarding the deep learning models, we used \textit{HuggingFace} for pretrained models whenever possible (\ie for AudioMAE, CLAP, DAC, MERT, and MuQ)\footnote{Respectively using the checkpoints {\small \texttt{hance-ai/audiomae}, \texttt{laion/clap-htsat-unfused}, \texttt{descript/dac\_44khz}, \texttt{MERT-v1-95M}}, and {\small \texttt{MuQ}~$|$~\texttt{MuQ-MuLan-large}} (for audio and multimodal spaces)}. All other pre-trained models 
were downloaded from their official repositories: CoDiCodec\footnote{\url{https://github.com/SonyCSLParis/codicodec}}, M2D\footnote{\url{https://github.com/nttcslab/m2d}}, MATPAC++\footnote{\url{https://github.com/aurianworld/matpac}}, and MusicFM\footnote{\url{https://github.com/minzwon/musicfm}} (MSD checkpoint). We release our code for reproducing experiments\footnote{\url{https://github.com/ax-le/msa_deep_embeddings}}.

\subsection{Hyperparameter Selection}
\label{sec:hyperparam_selection}
To ensure optimal and fair downstream performance, we performed a grid search over the hyperparameters of each segmentation algorithm. For Foote's algorithm, we searched the kernel size and the novelty curve median filtering over $\{8,12,16\}$. For LSD, we varied the number of clusters $k \in \{4,6,8,9,10,11,12,13,14,16\}$\footnote{We observe a high variability of the results given this parameter; detailed results are available in the Supplementary Material.} and set the median filtering parameter equal to $k$. 
For the CBM algorithm, we compared the Full and 7-band kernels. Crucially, we disabled the CBM penalty function that enforces specific segment sizes; omitting this penalty limits empirical structural priors and ensures the resulting segmentation better reflects the discriminative power of the evaluated embeddings. A study of the influence of these hyperparameters is presented in the Supplementary Material. LSD and Foote are computed using the \textit{MSAF}~\cite{msaf} toolbox, and the CBM using the original implementation.

This search yields, per model, a set of candidate configurations, from which we report two conditions. In the \emph{per-dataset} condition, the best configuration (averaging $\Fzf$ and $\Fth$) is selected per model \emph{and} dataset; this optimistic estimate reflects each embedding's intrinsic potential. In the \emph{cross-dataset} condition, a single configuration per model is selected, maximizing performance \emph{on average across all datasets}; this mimics a deployment setting where hyperparameters are fixed once and applied without re-tuning, though as the configuration still sees all datasets, it approximates rather than strictly tests generalization. Figures~\ref{fig:comp_deep_btf} and~\ref{fig:best_deep_models} use the \emph{per-dataset} condition, Table~\ref{tab:big_table} the \emph{cross-dataset} one.


\subsection{Metrics \& Datasets}
\label{sec:metrics}
Following standard protocols, we assess boundary detection using the hit-rate F-measure at 0.5s and 3s tolerances ($\Fzf$ and $\Fth$ respectively). Scores are computed using \textit{mir\_eval}~\cite{mireval}. We handle datasets via \textit{mirdata}~\cite{bittner2019mirdata} and evaluate our approach on three standard MSA benchmarks: RWC-Pop~\cite{rwc} (now open-source~\cite{balke2026rwc}), SALAMI~\cite{salami}, and Harmonix~\cite{nieto2019harmonix}. For RWC-Pop, we use the MIREX10 annotation set across its 100 popular music tracks. From the SALAMI dataset, we restrict our evaluation to the 884 tracks that possess two independent annotations as in~\cite{salamon2021deep,buisson2024using}; we evaluate against the coarse-level annotations and report the best score among both annotations. Finally, we utilize the 912 Western popular music tracks comprising the Harmonix set.

\subsection{Trimming}
\textit{Trimming} annotations and predictions consists of removing the first and last segments before computing scores. This practice remains largely overlooked in the MSA literature: to the best of our knowledge, only Buisson \etal~\cite{buisson2024using, buisson2024self} systematically trim annotations, and only the recent study of Korzeniowski and Vogl~\cite{korzeniowski2025simple} explicitly investigates the impact of trimming on segmentation results. Yet evaluating this impact is critical for a rigorous analysis: the absolute first and last boundaries carry no information about segmentation quality and can artificially inflate performance metrics.

In our experiments, we will present results \textbf{without trimming} first, to remain consistent with existing literature. Nonetheless, we will also present comparative results of our best scores with and without trimming to study the apparent drop in performance that occurs when this artificial inflation is removed.

\begin{figure*}[htb!]
     \centering
     \begin{subfigure}[b]{\textwidth}
         \centering
         \includegraphics[width=0.95\textwidth]{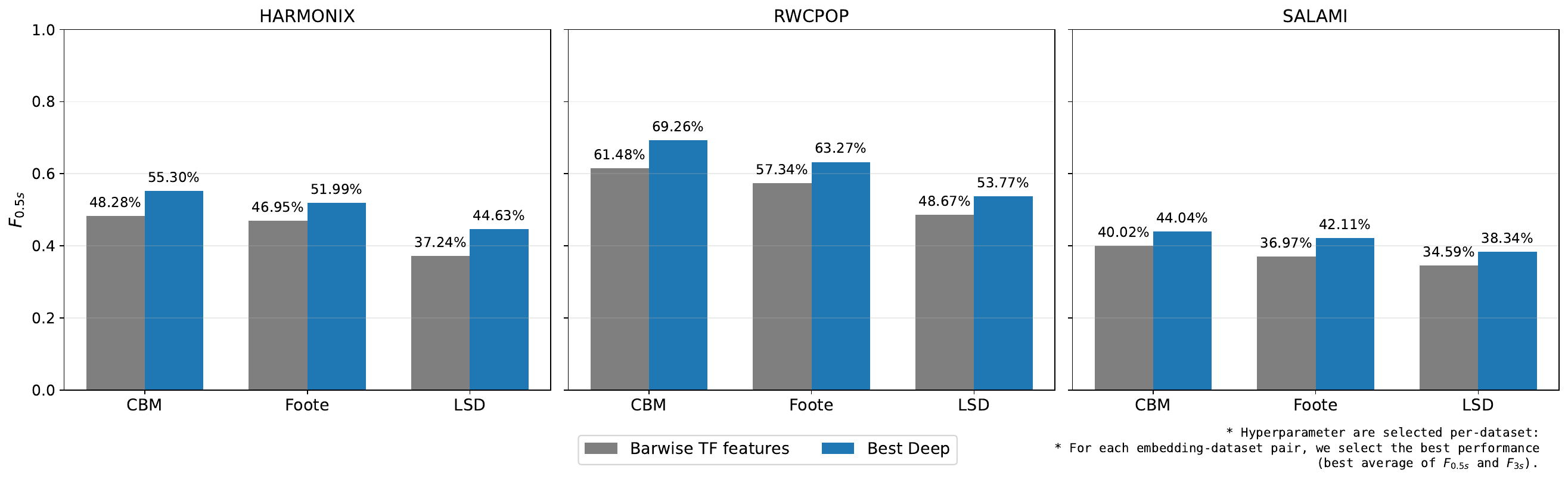}
         \caption{Best results obtained for $\Fzf$}
     \end{subfigure}
     \hfill

    \vspace{-1em}
    
     \centering
     \begin{subfigure}[b]{\textwidth}
         \centering
         \includegraphics[width=0.95\textwidth]{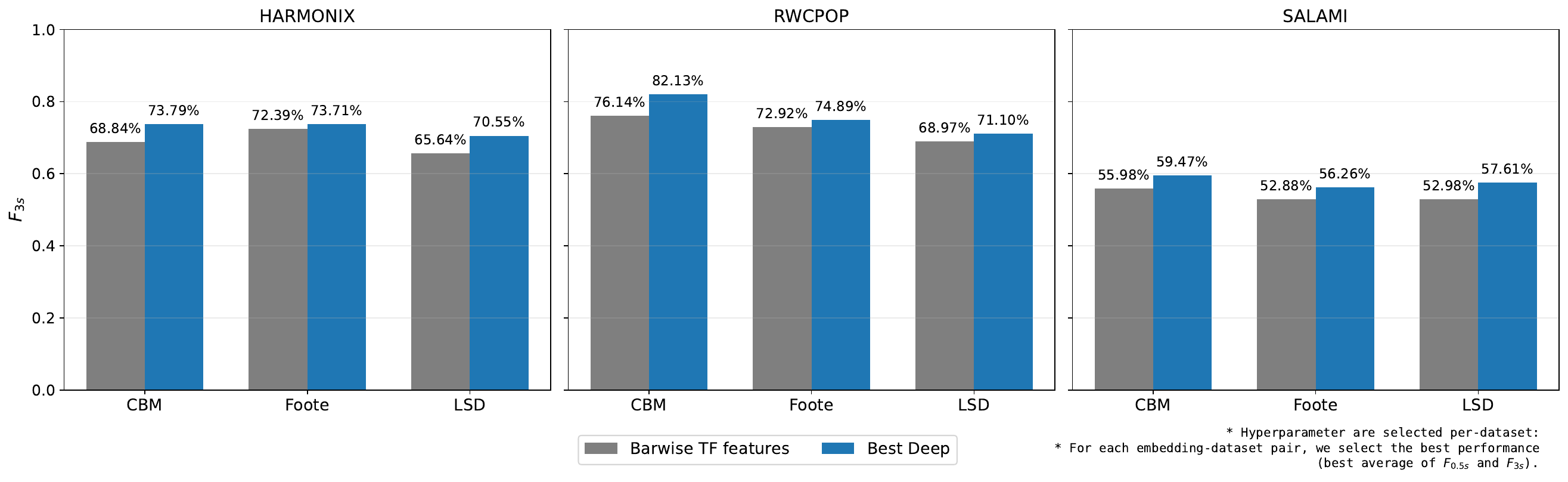}
         \caption{Best results obtained for $\Fth$}
     \end{subfigure}
     \hfill
    \vspace{-1.35em}
     \caption{Comparison of the best results obtained with deep models and the Barwise TF features (non-deep learning baseline), according to the segmentation algorithm and the dataset. The hyperparameters of the downstream segmentation algorithms are selected under the \emph{per-dataset} condition (Section~\ref{sec:hyperparam_selection}). 
     }
     \label{fig:comp_deep_btf}
\end{figure*}

\begin{figure*}[htb!]
    \centering  
    \includegraphics[width=0.9\textwidth]{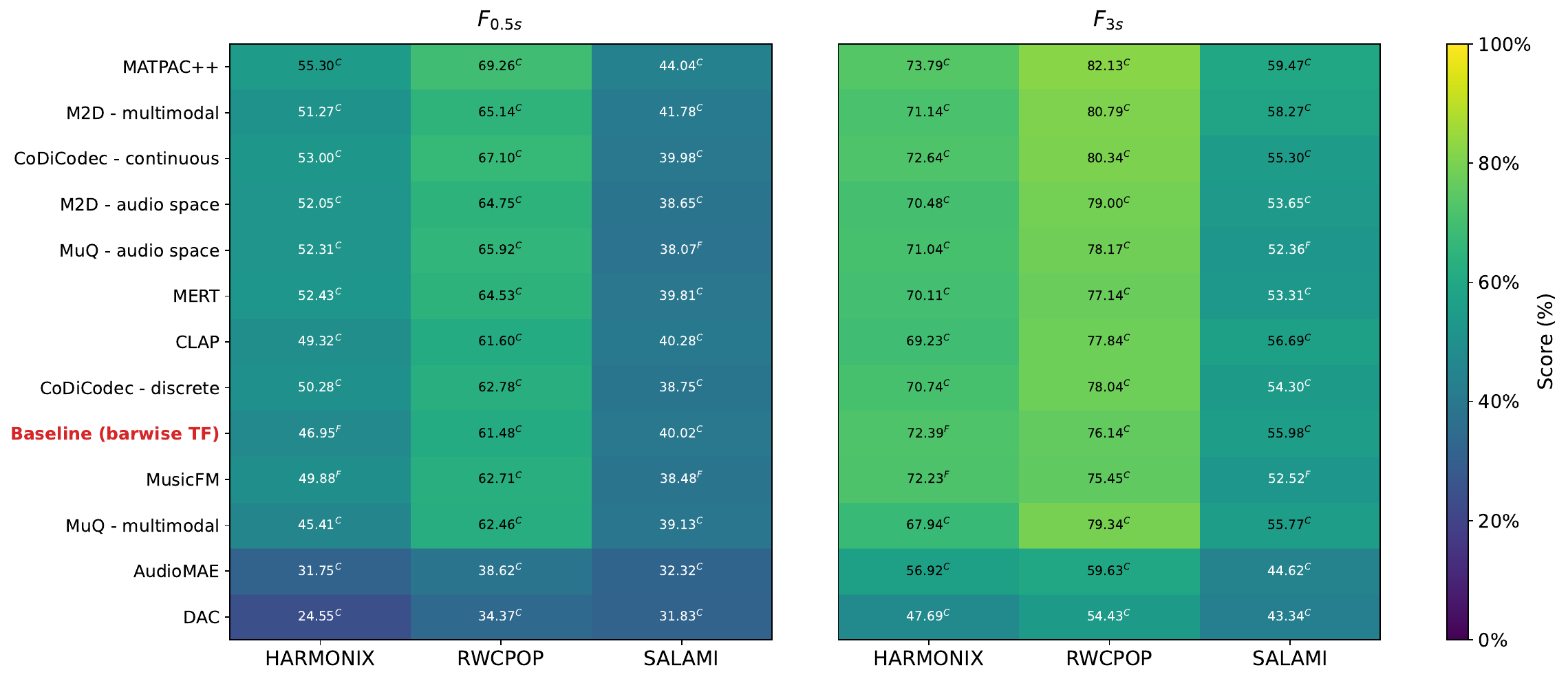}
        \vspace{-8pt}
    \caption{Best results obtained with all deep learning models, and their best downstream segmentation algorithm. Rows are ordered by decreasing average of $\Fzf$ and $\Fth$. Superscript denotes the downstream segmentation algorithm used to obtain these results ($C$: CBM, $F$: Foote). The hyperparameters of the downstream segmentation algorithms are selected under the \emph{per-dataset} condition (Section~\ref{sec:hyperparam_selection}).} 
    \vspace{-8pt}
    \label{fig:best_deep_models}
\end{figure*}

When analyzing the RWC-Pop (MIREX10) and SALAMI datasets, we observed that annotations systematically include silent segments at the file extremities (specifically, segments between 0 and the start of the audio signal, and between the end of the signal and the end of the file). We argue that these segments do not provide meaningful information regarding the quality of a structural prediction and should therefore be excluded. We term this preprocessing step ``double trimming'': first, removing the silent segments, and subsequently trimming both annotations and predictions to the active signal boundaries. We also evaluate the impact of this condition on the resulting performance scores. Since the annotations for the Harmonix dataset do not systematically contain these silences, we remove them by default; consequently, trimming on Harmonix is \textit{de facto} considered double trimming.

Ideally, we expect that future standards in MSA will adopt trimming, or even double trimming, by default.

\section{Results and Discussion}
\label{sec:results}

\begin{table*}[tb!]
\centering
\caption{Deep model segmentation performance across datasets (\%), using the CBM algorithm as the downstream segmentation algorithm (shown to be the best one on average). Hyperparameters are selected under the \emph{cross-dataset} condition (Section~\ref{sec:hyperparam_selection}). Best results per column are highlighted in \textbf{bold} (differentiating baselines and our methods). $^{\star}$represents models where annotations were trimmed. $^{\dagger}$represents results where the dataset was used for training (in cross-validation settings). $^{\ddagger}$represents oracle condition for hyperparameter fitting.}
\label{tab:big_table}
\begin{tabular}{l l c c c c c c}
\toprule
\multicolumn{2}{c}{\multirow{2}{*}{\textbf{Deep Model}}} & \multicolumn{2}{c}{\textbf{HARMONIX}} & \multicolumn{2}{c}{\textbf{RWCPOP}} & \multicolumn{2}{c}{\textbf{SALAMI}} \\
\cmidrule(lr){3-4} \cmidrule(lr){5-6} \cmidrule(lr){7-8}
\multicolumn{2}{c}{} & $\mathbf{F_{0.5}}$ & $\mathbf{F_3}$ & $\mathbf{F_{0.5}}$ & $\mathbf{F_3}$ & $\mathbf{F_{0.5}}$ & $\mathbf{F_3}$ \\
\midrule
\multicolumn{2}{l}{AudioMAE} & $31.75 \text{\scriptsize $\pm$ 18.21}$ & $56.92 \text{\scriptsize $\pm$ 18.46}$ & $38.62 \text{\scriptsize $\pm$ 16.19}$ & $59.63 \text{\scriptsize $\pm$ 18.01}$ & $25.86 \text{\scriptsize $\pm$ 12.98}$ & $43.51 \text{\scriptsize $\pm$ 15.76}$ \\
\cmidrule(lr){1-2} \cmidrule(lr){3-4} \cmidrule(lr){5-6} \cmidrule(lr){7-8}
\multicolumn{2}{l}{CLAP} & $49.32 \text{\scriptsize $\pm$ 21.38}$ & $69.23 \text{\scriptsize $\pm$ 17.26}$ & $61.60 \text{\scriptsize $\pm$ 17.98}$ & $77.84 \text{\scriptsize $\pm$ 17.01}$ & $35.18 \text{\scriptsize $\pm$ 16.70}$ & $51.72 \text{\scriptsize $\pm$ 17.32}$ \\
\cmidrule(lr){1-2} \cmidrule(lr){3-4} \cmidrule(lr){5-6} \cmidrule(lr){7-8}
\multirow{2}{*}{CoDiCodec} & Discrete & $50.28 \text{\scriptsize $\pm$ 20.52}$ & $70.74 \text{\scriptsize $\pm$ 16.02}$ & $62.78 \text{\scriptsize $\pm$ 16.98}$ & $78.04 \text{\scriptsize $\pm$ 15.27}$ & $35.01 \text{\scriptsize $\pm$ 16.74}$ & $50.51 \text{\scriptsize $\pm$ 17.72}$ \\
\cmidrule(lr){2-2} \cmidrule(lr){3-4} \cmidrule(lr){5-6} \cmidrule(lr){7-8}
& Continuous & $53.00 \text{\scriptsize $\pm$ 20.70}$ & $72.64 \text{\scriptsize $\pm$ 16.34}$ & $67.10 \text{\scriptsize $\pm$ 18.25}$ & $80.34 \text{\scriptsize $\pm$ 15.11}$ & $36.54 \text{\scriptsize $\pm$ 17.31}$ & $51.73 \text{\scriptsize $\pm$ 17.55}$ \\
\cmidrule(lr){1-2} \cmidrule(lr){3-4} \cmidrule(lr){5-6} \cmidrule(lr){7-8}
\multicolumn{2}{l}{DAC} & $24.10 \text{\scriptsize $\pm$ 19.38}$ & $47.08 \text{\scriptsize $\pm$ 21.57}$ & $34.37 \text{\scriptsize $\pm$ 16.83}$ & $54.43 \text{\scriptsize $\pm$ 19.45}$ & $25.59 \text{\scriptsize $\pm$ 13.17}$ & $42.87 \text{\scriptsize $\pm$ 15.71}$ \\
\cmidrule(lr){1-2} \cmidrule(lr){3-4} \cmidrule(lr){5-6} \cmidrule(lr){7-8}
\multirow{2}{*}{M2D} & Audio space & $52.05 \text{\scriptsize $\pm$ 18.79}$ & $70.48 \text{\scriptsize $\pm$ 14.30}$ & $65.14 \text{\scriptsize $\pm$ 13.77}$ & $75.79 \text{\scriptsize $\pm$ 12.70}$ & $36.02 \text{\scriptsize $\pm$ 16.88}$ & $49.29 \text{\scriptsize $\pm$ 17.95}$ \\
\cmidrule(lr){2-2} \cmidrule(lr){3-4} \cmidrule(lr){5-6} \cmidrule(lr){7-8}
& Multimodal & $50.90 \text{\scriptsize $\pm$ 21.50}$ & $70.59 \text{\scriptsize $\pm$ 17.22}$ & $65.14 \text{\scriptsize $\pm$ 15.97}$ & $80.79 \text{\scriptsize $\pm$ 15.42}$ & $36.10 \text{\scriptsize $\pm$ 17.24}$ & $52.58 \text{\scriptsize $\pm$ 17.71}$ \\
\cmidrule(lr){1-2} \cmidrule(lr){3-4} \cmidrule(lr){5-6} \cmidrule(lr){7-8}
\multicolumn{2}{l}{MATPAC++} & $\mathbf{55.30 \text{\scriptsize $\pm$ 20.83}}$ & $\mathbf{73.79 \text{\scriptsize $\pm$ 15.75}}$ & $\mathbf{69.26 \text{\scriptsize $\pm$ 16.42}}$ & $\mathbf{82.13 \text{\scriptsize $\pm$ 14.88}}$ & $38.90 \text{\scriptsize $\pm$ 17.66}$ & $53.74 \text{\scriptsize $\pm$ 17.74}$ \\
\cmidrule(lr){1-2} \cmidrule(lr){3-4} \cmidrule(lr){5-6} \cmidrule(lr){7-8}
\multicolumn{2}{l}{MERT} & $52.43 \text{\scriptsize $\pm$ 19.25}$ & $70.11 \text{\scriptsize $\pm$ 14.98}$ & $64.53 \text{\scriptsize $\pm$ 15.41}$ & $77.14 \text{\scriptsize $\pm$ 13.79}$ & $36.81 \text{\scriptsize $\pm$ 16.80}$ & $49.52 \text{\scriptsize $\pm$ 17.98}$ \\
\cmidrule(lr){1-2} \cmidrule(lr){3-4} \cmidrule(lr){5-6} \cmidrule(lr){7-8}
\multirow{2}{*}{MuQ} & Audio space & $51.97 \text{\scriptsize $\pm$ 18.66}$ & $69.67 \text{\scriptsize $\pm$ 14.62}$ & $65.92 \text{\scriptsize $\pm$ 13.99}$ & $78.17 \text{\scriptsize $\pm$ 13.06}$ & $35.68 \text{\scriptsize $\pm$ 16.89}$ & $48.73 \text{\scriptsize $\pm$ 17.70}$ \\
\cmidrule(lr){2-2} \cmidrule(lr){3-4} \cmidrule(lr){5-6} \cmidrule(lr){7-8}
& Multimodal & $45.41 \text{\scriptsize $\pm$ 20.40}$ & $67.94 \text{\scriptsize $\pm$ 17.27}$ & $62.46 \text{\scriptsize $\pm$ 16.57}$ & $79.34 \text{\scriptsize $\pm$ 14.93}$ & $33.95 \text{\scriptsize $\pm$ 15.73}$ & $51.30 \text{\scriptsize $\pm$ 16.97}$ \\
\cmidrule(lr){1-2} \cmidrule(lr){3-4} \cmidrule(lr){5-6} \cmidrule(lr){7-8}
\multicolumn{2}{l}{MusicFM} & $51.55 \text{\scriptsize $\pm$ 18.18}$ & $70.52 \text{\scriptsize $\pm$ 13.86}$ & $59.66 \text{\scriptsize $\pm$ 11.58}$ & $71.50 \text{\scriptsize $\pm$ 11.35}$ & $38.37 \text{\scriptsize $\pm$ 16.02}$ & $51.95 \text{\scriptsize $\pm$ 16.88}$ \\

\midrule

\multicolumn{8}{c}{\textbf{Barwise TF baselines}} \\
\midrule
\multicolumn{2}{l}{Foote} & $46.95 \text{\scriptsize $\pm$ 19.18}$ & $72.39 \text{\scriptsize $\pm$ 14.63}$ & $57.50 \text{\scriptsize $\pm$ 13.70}$ & $71.41 \text{\scriptsize $\pm$ 13.72}$ & $36.96 \text{\scriptsize $\pm$ 15.67}$ & $52.87 \text{\scriptsize $\pm$ 16.98}$ \\
\cmidrule(lr){1-2} \cmidrule(lr){3-4} \cmidrule(lr){5-6} \cmidrule(lr){7-8}
\multicolumn{2}{l}{LSD} & $37.41 \text{\scriptsize $\pm$ 16.07}$ & $65.21 \text{\scriptsize $\pm$ 14.24}$ & $48.17 \text{\scriptsize $\pm$ 12.73}$ & $67.48 \text{\scriptsize $\pm$ 14.68}$ & $33.70 \text{\scriptsize $\pm$ 12.61}$ & $52.51 \text{\scriptsize $\pm$ 14.65}$ \\
\cmidrule(lr){1-2} \cmidrule(lr){3-4} \cmidrule(lr){5-6} \cmidrule(lr){7-8}
\multicolumn{2}{l}{CBM} & $48.28 \text{\scriptsize $\pm$ 17.56}$ & $68.84 \text{\scriptsize $\pm$ 13.25}$ & $58.51 \text{\scriptsize $\pm$ 12.60}$ & $71.27 \text{\scriptsize $\pm$ 11.71}$ & $\mathbf{40.02 \text{\scriptsize $\pm$ 14.76}}$ & $\mathbf{55.98 \text{\scriptsize $\pm$ 15.16}}$ \\

\midrule

\multicolumn{8}{c}{\textbf{Literature results}} \\
\midrule
\multicolumn{2}{l}{AudioMAE scores from~\cite{toyama2026foundational}}  & 36.95 & 58.11 & - & - & - & - \\
\cmidrule(lr){1-2} \cmidrule(lr){3-4} \cmidrule(lr){5-6} \cmidrule(lr){7-8}
\multicolumn{2}{l}{CLAP scores from~\cite{toyama2026foundational}}  & 29.21 & 46.60 & - & - & - & - \\
\cmidrule(lr){1-2} \cmidrule(lr){3-4} \cmidrule(lr){5-6} \cmidrule(lr){7-8}
\multicolumn{2}{l}{DAC scores from~\cite{toyama2026foundational}}  & 19.10 & 39.63 & - & - & - & - \\
\cmidrule(lr){1-2} \cmidrule(lr){3-4} \cmidrule(lr){5-6} \cmidrule(lr){7-8}
\multicolumn{2}{l}{MERT scores from~\cite{toyama2026foundational}}  & 42.23 & 60.99 & - & - & - & - \\
\cmidrule(lr){1-2} \cmidrule(lr){3-4} \cmidrule(lr){5-6} \cmidrule(lr){7-8}
\multicolumn{2}{l}{MusicFM scores from~\cite{toyama2026foundational}}  & 49.76 & 63.91 & - & - & - & - \\
\cmidrule(lr){1-2} \cmidrule(lr){3-4} \cmidrule(lr){5-6} \cmidrule(lr){7-8}
\multicolumn{2}{l}{Salamon \etal~\cite{salamon2021deep}}  & 45.74 & 68.84 & - & - & 33.78 & 55.65 \\
\cmidrule(lr){1-2} \cmidrule(lr){3-4} \cmidrule(lr){5-6} \cmidrule(lr){7-8}
\multicolumn{2}{l}{Wang \etal~\cite{wang2021supervised}}  & 49.7 & 73.8 & - & - & - & - \\
\cmidrule(lr){1-2} \cmidrule(lr){3-4} \cmidrule(lr){5-6} \cmidrule(lr){7-8}
\multicolumn{2}{l}{Buisson \etal~(supervised)~\cite{buisson2024using}{$^{\star}$}}  & 56.8$^{\dagger}$ & 71.7$^{\dagger}$ & 58.5$^{\dagger}$ & 75.0$^{\dagger}$ & - & - \\
\cmidrule(lr){1-2} \cmidrule(lr){3-4} \cmidrule(lr){5-6} \cmidrule(lr){7-8}
\multirow{2}{*}{\allinone~\cite{kim2023all}} & from~\cite{kim2023all} & \textbf{66.0}$^{\dagger}$ & - & - & - & - & - \\
& run locally & \multicolumn{2}{c}{{\scriptsize(Not reported - train/test leakage)}} & $\mathbf{71.10} \text{\scriptsize $\pm$ 14.28}$ & $\mathbf{80.07} \text{\scriptsize $\pm$ 11.01}$ & $\mathbf{47.75} \text{\scriptsize $\pm$ 16.62}$ & $60.06 \text{\scriptsize $\pm$ 15.30}$ \\
\cmidrule(lr){1-2} \cmidrule(lr){3-4} \cmidrule(lr){5-6} \cmidrule(lr){7-8}
\multicolumn{2}{l}{Buisson \etal~(SSL)~\cite{buisson2024self}$^{\star \ddagger}$}  & 48.5 & \textbf{80.8} & - & - & 39.8 \text{\scriptsize$\pm$ 16} & \textbf{70.1} \text{\scriptsize $\pm$17} \\
\bottomrule
\end{tabular}
\vspace{-7pt}
\end{table*}

\subsection{Deep Models \textit{vs.} Barwise TF features}

Figure~\ref{fig:comp_deep_btf} compares peak deep audio models' performance against the Barwise TF baseline. Deep embeddings achieve the highest scores across all datasets and algorithms, demonstrating clear improvements over spectrogram-based features. 

Figure~\ref{fig:best_deep_models} ranks the nine evaluated deep models. Notably, three of them (and one multimodal space) are outperformed by the baseline. The fact that nearly a third of the models fail to beat traditional acoustic features suggests that their specific pretraining objectives or latent spaces may not align well with structural musical properties. Among the successful models, MATPAC++ achieves the highest overall performance, with M2D and the continuous space of CoDiCodec closely following, indicating their latent spaces are particularly well-suited to disambiguate barwise audio representations. Although we observe differences when using multimodal versus unimodal audio spaces, the results are contradictory across conditions and yield no conclusive advantage for either approach. Both discrete embedding spaces (DAC and the discrete version of CoDiCodec) yielded relatively poor performance, with CoDiCodec's discrete representation notably underperforming its continuous counterpart. This suggests that discrete embeddings are ill-suited for our methodology. Finally, models trained exclusively on music data (MERT, MusicFM, and MuQ) consistently underperform, with results ranging from medium to poor.

Overall, these results demonstrate that while the rich semantic and temporal contexts of learned representations provide a distinct advantage for boundary detection, off-the-shelf deep audio embeddings do not universally guarantee improved segmentation over standard feature-based representations.
\begin{table*}
\caption{MATPAC++, Barwise TF-CBM, and \allinone results from~\tabref{tab:big_table}, in different trimming conditions.}
\vspace{-2pt}
\label{tab:trimmed_res}
\begin{tabular}{llcccccc}
\toprule
\multicolumn{2}{c}{\multirow{2}{*}{\textbf{Method}}} & \multicolumn{2}{c}{\textbf{HARMONIX}} & \multicolumn{2}{c}{\textbf{RWCPOP}} & \multicolumn{2}{c}{\textbf{SALAMI}} \\
\cmidrule(lr){3-4} \cmidrule(lr){5-6} \cmidrule(lr){7-8}
\multicolumn{2}{c}{} & $\mathbf{F_{0.5}}$ & $\mathbf{F_3}$ & $\mathbf{F_{0.5}}$ & $\mathbf{F_3}$ & $\mathbf{F_{0.5}}$ & $\mathbf{F_3}$ \\
\midrule
\multirow{3}{*}{MATPAC++} & No trimming & $55.30 \text{\scriptsize $\pm$ 20.83}$ & $73.79 \text{\scriptsize $\pm$ 15.75}$ & $69.26 \text{\scriptsize $\pm$ 16.42}$ & $82.13 \text{\scriptsize $\pm$ 14.88}$ & $38.90 \text{\scriptsize $\pm$ 17.66}$ & $53.74 \text{\scriptsize $\pm$ 17.74}$ \\
\cmidrule(lr){2-2} \cmidrule(lr){3-4} \cmidrule(lr){5-6} \cmidrule(lr){7-8}
& Trimming & $-$ & $-$ & $65.39 \text{\scriptsize $\pm$ 18.48}$ & $79.87 \text{\scriptsize $\pm$ 16.83}$ & $30.05 \text{\scriptsize $\pm$ 19.51}$ & $47.12 \text{\scriptsize $\pm$ 19.70}$ \\
\cmidrule(lr){2-2} \cmidrule(lr){3-4} \cmidrule(lr){5-6} \cmidrule(lr){7-8}
& Double trimming & $54.27 \text{\scriptsize $\pm$ 23.35}$ & $70.96 \text{\scriptsize $\pm$ 17.83}$ & $65.04 \text{\scriptsize $\pm$ 21.67}$ & $77.88 \text{\scriptsize $\pm$ 19.10}$ & $28.20 \text{\scriptsize $\pm$ 22.21}$ & $43.33 \text{\scriptsize $\pm$ 21.93}$ \\
\midrule
Barwise TF & No trimming & $48.28 \text{\scriptsize $\pm$ 17.56}$ & $68.84 \text{\scriptsize $\pm$ 13.25}$ & $58.51 \text{\scriptsize $\pm$ 12.60}$ & $71.27 \text{\scriptsize $\pm$ 11.71}$ & $40.02 \text{\scriptsize $\pm$ 14.76}$ & $55.98 \text{\scriptsize $\pm$ 15.16}$ \\
\cmidrule(lr){2-2} \cmidrule(lr){3-4} \cmidrule(lr){5-6} \cmidrule(lr){7-8}
CBM & Trimming & $-$ & $-$ & $53.18 \text{\scriptsize $\pm$ 14.33}$ & $67.59 \text{\scriptsize $\pm$ 13.26}$ & $28.80 \text{\scriptsize $\pm$ 16.78}$ & $47.65 \text{\scriptsize $\pm$ 17.98}$ \\
\cmidrule(lr){2-2} \cmidrule(lr){3-4} \cmidrule(lr){5-6} \cmidrule(lr){7-8}
 & Double trimming & $45.55 \text{\scriptsize $\pm$ 19.79}$ & $65.10 \text{\scriptsize $\pm$ 15.14}$ & $51.00 \text{\scriptsize $\pm$ 16.67}$ & $64.08 \text{\scriptsize $\pm$ 15.09}$ & $25.68 \text{\scriptsize $\pm$ 19.78}$ & $42.38 \text{\scriptsize $\pm$ 21.67}$ \\
\midrule
\multirow{3}{*}{\allinone \cite{kim2023all}} & No trimming & \multicolumn{2}{c}{{\scriptsize(Not reported - train/test leakage)}} & $71.10 \text{\scriptsize $\pm$ 14.28}$ & $80.07 \text{\scriptsize $\pm$ 11.01}$ & $47.75 \text{\scriptsize $\pm$ 16.62}$ & $60.06 \text{\scriptsize $\pm$ 15.30}$ \\
\cmidrule(lr){2-2} \cmidrule(lr){3-4} \cmidrule(lr){5-6} \cmidrule(lr){7-8}

 & Trimming & $-$ & $-$ & $67.13 \text{\scriptsize $\pm$ 16.22}$ & $77.27  \text{\scriptsize $\pm$ 12.59}$ & $37.50 \text{\scriptsize $\pm$ 19.46}$ & $52.23 \text{\scriptsize $\pm$ 18.00}$ \\
\cmidrule(lr){2-2} \cmidrule(lr){3-4} \cmidrule(lr){5-6} \cmidrule(lr){7-8}
 & Double trimming & $-$ & $-$ & $65.30 \text{\scriptsize $\pm$ 18.00}$ & $74.61 \text{\scriptsize $\pm$ 13.89}$ & $32.70 \text{\scriptsize $\pm$ 24.03}$ & $46.73 \text{\scriptsize $\pm$ 22.74}$ \\
\bottomrule
\end{tabular}
\vspace{-7pt}
\end{table*}

\subsection{Downstream Segmentation Algorithm}
As indicated by the results in Figure~\ref{fig:best_deep_models}, CBM is the most effective downstream segmentation algorithm for the vast majority of embedding-dataset pairs. Although the Foote algorithm occasionally achieves superior performance (notably with CoDiCodec), CBM performs best in 33 out of 36 conditions. These results demonstrate the interest of the CBM in MSA, even in conjunction with deep embeddings.

This study also marks the first application of the LSD algorithm at the bar scale. Unlike the Foote and CBM algorithms, designed to identify novelty and homogeneity (\ie~abrupt changes or local similarity), LSD is specialized for detecting "stripes" (\ie~repeating patterns). Given its proven effectiveness in the literature~\cite{buisson2024self}, we hypothesize that LSD's relative underperformance here stems from a misalignment with the barwise representations (a premise for barwise processing is a higher homogeneity between bars than beats) rather than a limitation of the algorithm itself to segment deep embeddings.

\subsection{Results compared with literature}

Next, we compare our approach against state-of-the-art models in~\tabref{tab:big_table}. For this comparison, we report results using only the CBM downstream algorithm, under the \emph{cross-dataset} condition for hyperparameters (Section~\ref{sec:hyperparam_selection}). We also report results obtained by running locally the \allinone~\cite{kim2023all} model, on RWC-Pop and SALAMI only, since Harmonix is part of its training data.

Five of the nine models (AudioMAE~\cite{huang2022masked}, MusicFM~\cite{won2024foundation}, MERT~\cite{MERT}, DAC~\cite{kumar2023high}, and CLAP~\cite{CLAP}) were compared in~\cite{toyama2026foundational}, where their embeddings were studied in a linear probing fashion. Comparing our CBM-based downstream approach to these linear probing baselines on the Harmonix dataset, we observe that our methodology yields significantly stronger performance for almost all shared models. Most notably, CLAP and MERT show substantial improvements, rising from 29.21\% to 49.32\% and from 42.23\% to 52.43\% for $\Fzf$, respectively. This suggests that the CBM algorithm and barwise processing are highly effective at extracting structural information from these deep audio embeddings, notably when compared with linear probing.

Overall, three methods stand out: the self-supervised approach of Buisson \etal~\cite{buisson2024self}, obtaining the best $\Fth$ results by a wide margin; the supervised \allinone, obtaining the best $\Fzf$ results on the datasets it did not train on; and, among training-free generic embeddings, MATPAC++. In particular, we observe that MATPAC++ proves highly competitive with specialized literature on both Harmonix and RWC-Pop: only the two models from Buisson \etal~\cite{buisson2024using,buisson2024self} and \allinone~\cite{kim2023all} surpass its performance, while it required no supervision. In addition, the performance on the RWC-Pop dataset is on par with the best-performing model (\allinone).

Focusing on Buisson \etal, we stress that their approach and ours differ methodologically: they train a model \emph{purpose-built} for MSA, with structure-specific inductive biases, whereas MATPAC++ is a \emph{generic} audio model, simply paired with a standard, unsupervised segmentation algorithm. Their scores also benefit from a favorable evaluation protocol: their model produces a hierarchy of segmentations (by varying the number of clusters in LSD), and only the level maximizing each metric is reported per track~\cite{buisson2024self}. This oracle selection is a legitimate and explicitly stated choice, but one that our single-level protocol does not exploit, suggesting the true gap is even smaller than the raw scores indicate.

\newpage
\allinone, in contrast, was used out-of-the-box, without modification or favorable tuning. Its performance comes at a cost, however: it is fully supervised and relies on rich, task-specific inputs, including source-separated instrument stems. MATPAC++ uses none of this, yet trails \allinone~by only a few points on RWC-Pop. That MATPAC++, a generic, training-free embedding using none of this, stays within reach of both a purpose-built self-supervised model and a fully supervised state-of-the-art system is, in our view, the central finding: general-purpose representations already encode rich structural information.

We conclude that while SSL representations dedicated to MSA still offer significant benefits, there is room for refinement. Specifically, we hypothesize that incorporating barwise processing could substantially improve $\Fzf$ performance. Developing future bar-scale SSL models, segmented with a standard algorithm (notably CBM), may thus improve current results.

\subsection{Trimming results}
Finally, we present trimmed results in~\tabref{tab:trimmed_res}. These results demonstrate a consistent loss of performance when trimming annotations, which was expected and consistent with~\cite{korzeniowski2025simple}. While trimming has only a marginal impact on the Harmonix dataset, the degradation can be severe elsewhere; for instance, $\Fzf$ performance drops by more than $11\%$ for the Barwise TF CBM on the SALAMI dataset, and degrades even further with double trimming. Because standard evaluation without trimming can artificially inflate scores by rewarding trivial boundary matches, we advocate that trimming should be adopted as the new standard for rigorous MSA evaluation.

\section{Conclusion}
\label{sec:conclusion}

In this work, we investigated the training-free capabilities of nine generic deep audio models for Music Structure Analysis, specifically focusing on boundary retrieval. By leveraging unsupervised downstream segmentation algorithms at the bar scale, we demonstrated that deep audio embeddings generally provide a distinct advantage over traditional barwise acoustic features. Among the evaluated representations, MATPAC++ proved particularly well-suited for estimating structural boundaries. Furthermore, our comparative analysis established the Correlation Block-Matching algorithm as the most effective downstream segmentation method, notably outperforming recent linear probing approaches.

Despite these advancements, the strongest results are still obtained by models specifically designed for the task: a self-supervised and a fully supervised approach. Our contribution is therefore not to surpass these specialized systems, but to show that generic, off-the-shelf representations reach competitive performance at no task-specific training cost. Future work could bring the effective ingredients identified here (bar-scale processing and standard segmentation algorithms such as CBM) to specialized, notably SSL, models, which could help close the remaining gap. Additionally, we demonstrated that traditional evaluation metrics are often artificially inflated by initial and final boundary matches. We strongly advocate for the community to adopt standard trimming and double trimming practices for more rigorous future evaluations.

\section{acknowledgments}
With support of ANR JCJC MusAIc ANR-25-CE23-5316.

\bibliography{smc2026bib}

@inproceedings{mccallum2019unsupervised,
  title={Unsupervised learning of deep features for music segmentation},
  author={McCallum, Matthew C},
  booktitle={Int. Conf. Acoustics, Speech and Signal Processing (ICASSP)},
  year={2019},
  organization={IEEE}
}

@article{serra2014unsupervised,
  title={Unsupervised music structure annotation by time series structure features and segment similarity},
  author={Serr{\`a}, Joan and M{\"u}ller, Meinard and Grosche, Peter and Arcos, Josep Ll},
  journal={IEEE Trans. Multimedia},
  volume={16},
  number={5},
  year={2014},
  publisher={IEEE}
}

@inproceedings{wang2021supervised,
  title={Supervised Metric Learning for Music Structure Feature},
  author={Wang, Ju-Chiang and Smith, Jordan BL and Lu, Wei-Tsung and Song, Xuchen},
  booktitle={Int. Soc. Music Information Retrieval Conf. (ISMIR)},
  year={2021},
}

@inproceedings{grill2015cnn,
  title={Music Boundary Detection Using Neural Networks on Combined Features and Two-Level Annotations},
  author={Grill, Thomas and Schl{\"u}ter, Jan},
  booktitle={Int. Soc. Music Information Retrieval Conf. (ISMIR)},
  year={2015}
}

@inproceedings{nieto2019harmonix,
  title={The {Harmonix} Set: Beats, Downbeats, and Functional Segment Annotations of Western Popular Music},
  author={Nieto, Oriol and others},
  booktitle={Int. Soc. Music Information Retrieval Conf. (ISMIR)},
  year={2019}
}

@inproceedings{salami,
  title={Design and creation of a large-scale database of structural annotations},
  author={Smith, Jordan BL and others},
  booktitle={Int. Soc. Music Information Retrieval Conf. (ISMIR)},
  year={2011}}

@inproceedings{foote2000automatic,
  title={Automatic audio segmentation using a measure of audio novelty},
  author={Foote, Jonathan},
  booktitle={Int. Conf. Multimedia and Expo},
  year={2000},
  organization={IEEE}
}

@inproceedings{mcfee2014analyzing,
  title={Analyzing Song Structure with Spectral Clustering},
  author={McFee, Brian and Ellis, Dan},
  booktitle={Int. Soc. Music Information Retrieval Conf. (ISMIR)},
  year={2014}
}

@article{nieto2020segmentationreview,
  title={Audio-Based Music Structure Analysis: Current Trends, Open Challenges, and Applications},
  author={Nieto, Oriol and others},
  journal={Trans. Int. Soc. Music Information Retrieval},
  volume={3},
  number={1},
  year={2020},
  publisher={Ubiquity Press}
}

@inproceedings{salamon2021deep,
  title={Deep Embeddings and Section Fusion Improve Music Segmentation},
  author={Salamon, Justin and Nieto, Oriol and Bryan, Nicholas J},
  booktitle={Int. Soc. Music Information Retrieval Conf. (ISMIR)},
  year={2021}
}

@inproceedings{rwc,
  title={{RWC Music Database: Popular, Classical and Jazz Music Databases}},
  author={Goto, Masataka and Hashiguchi, Hiroki and Nishimura, Takuichi and Oka, Ryuichi},
  booktitle={Int. Soc. Music Information Retrieval Conf. (ISMIR)},
  year={2002}
}

@inproceedings{msaf,
  title={Systematic exploration of computational music structure research},
  author={Nieto, Oriol and Bello, Juan Pablo},
  booktitle={Int. Soc. Music Information Retrieval Conf. (ISMIR)},
  year={2016}
}

@inproceedings{mireval,
  title={mir\_eval: A transparent implementation of common {MIR} metrics},
  author={Raffel, Colin and others},
  booktitle={Int. Soc. Music Information Retrieval Conf. (ISMIR)},
  year={2014},
}

@article{marmoret2023barwise,
  title={Barwise Music Structure Analysis with the Correlation Block-Matching Segmentation Algorithm},
  author={Marmoret, Axel and Cohen, J{\'e}r{\'e}my E and Bimbot, Fr{\'e}d{\'e}ric},
  journal={Trans. Int. Soc. Music Information Retrieval},
  volume={6},
  number={1},
  year={2023}
}

@article{morreale2025reductive,
  title={Reductive, exclusionary, normalising: the limits of generative AI music},
  author={Morreale, Fabio and Martinez-Ramirez, Marco A and Masu, Raul and Liao, WeiHsiang and Mitsufuji, Yuki},
  journal={Trans. Int. Soc. Music Information Retrieval},
  volume={8},
  number={1},
  year={2025}
}

@inproceedings{choi2025understanding,
  title={Understanding the Potentials and Limitations of Prompt-based Music Generative AI},
  author={Choi, Youjin and Moon, JaeYoung and Yoo, JinYoung and Hong, Jin-Hyuk},
  booktitle={Proc. 2025 CHI Conf. on Human Factors in Computing Systems},
  year={2025}
}

@inproceedings{toyama2026foundational,
  title={Do Foundational Audio Encoders Understand Music Structure?},
  author={Toyama, Keisuke and others},
  booktitle={Int. Conf. Acoustics, Speech and Signal Processing (ICASSP)},
  year={2026}
}

@inproceedings{buisson2024using,
  title={Using pairwise link prediction and graph attention networks for music structure analysis},
  author={Buisson, Morgan and Mcfee, Brian and Essid, Slim},
  booktitle={Int. Soc. Music Information Retrieval Conf. (ISMIR)},
  year={2024}
}

@inproceedings{foscarin2024beat,
  title={{Beat This!} accurate beat tracking without dbn postprocessing},
  author={Foscarin, Francesco and Schl{\"u}ter, Jan and Widmer, Gerhard},
  booktitle={Int. Soc. Music Information Retrieval Conf. (ISMIR)},
  year={2024}
}

@article{buisson2024self,
  title={Self-supervised learning of multi-level audio representations for music segmentation},
  author={Buisson, Morgan and McFee, Brian and Essid, Slim and Crayencour, H{\'e}l{\`e}ne C},
  journal={IEEE/ACM Trans. Audio, Speech and Language Processing},
  volume={32},
  year={2024},
  publisher={IEEE}
}

@inproceedings{pasini2025codicodec,
  title={{CoDiCodec}: Unifying Continuous and Discrete Compressed Representations of Audio},
  author={Pasini, Marco and Lattner, Stefan and Fazekas, George},
  booktitle={Int. Soc. Music Information Retrieval Conf. (ISMIR)},
  year={2025}
}

@article{kumar2023high,
  title={High-fidelity audio compression with improved {RVQGAN}},
  author={Kumar, Rithesh and Seetharaman, Prem and Luebs, Alejandro and Kumar, Ishaan and Kumar, Kundan},
  journal={Advances in Neural Information Processing Systems},
  volume={36},
  year={2023}
}

@inproceedings{niizumi2024m2d,
  title={{M2D-CLAP}: Masked Modeling Duo Meets CLAP for Learning General-purpose Audio-Language Representation},
  author={Niizumi, Daisuke and others},
  booktitle={Interspeech},
  year={2024},
  publisher={ISCA}
}

@inproceedings{huang2022masked,
  title={Masked autoencoders that listen},
  author={Huang, Po-Yao and others},
  booktitle={Advances in Neural Information Processing Systems},
  volume={35},
  year={2022}
}

@article{quelennec2025matpac_plus,
  title={{MATPAC++}: Enhanced Masked Latent Prediction for Self-Supervised Audio Representation Learning},
  author={Quelennec, Aurian and Chouteau, Pierre and Peeters, Geoffroy and Essid, Slim},
  journal={arXiv preprint arXiv:2508.12709},
  year={2025}
}

@inproceedings{won2024foundation,
  title={A foundation model for music informatics},
  author={Won, Minz and Hung, Yun-Ning and Le, Duc},
  booktitle={Int. Conf. Acoustics, Speech and Signal Processing (ICASSP)},
  year={2024},
  organization={IEEE}
}

@article{zhu2025muq,
  title={{MuQ}: Self-supervised music representation learning with mel residual vector quantization},
  author={Zhu, Haina and others},
  journal={IEEE/ACM Trans. Audio, Speech and Language Processing},
  year={2025},
  publisher={IEEE}
}

@article{MERT,
  title={{MERT}: Acoustic music understanding model with large-scale self-supervised training},
  author={Li, Yizhi and others},
  journal={arXiv preprint arXiv:2306.00107},
  year={2023}
}

@inproceedings{CLAP,
  title={Large-scale contrastive language-audio pretraining with feature fusion and keyword-to-caption augmentation},
  author={Wu, Yusong and others},
  booktitle={Int. Conf. Acoustics, Speech and Signal Processing (ICASSP)},
  year={2023},
  organization={IEEE}
}

@inproceedings{kim2023all,
  title={All-in-one metrical and functional structure analysis with neighborhood attentions on demixed audio},
  author={Kim, Taejun and Nam, Juhan},
  booktitle={Workshop on Applications of Signal Processing to Audio and Acoustics (WASPAA)},
  year={2023},
  organization={IEEE}
}

@inproceedings{bittner2019mirdata,
  title={mirdata: Software for Reproducible Usage of Datasets.},
  author={Bittner, Rachel M and others},
  booktitle={Int. Soc. Music Information Retrieval Conf. (ISMIR)},
  year={2019}
}

@article{balke2026rwc,
  title={{RWC} Revisited: Towards a Community-Driven {MIR} Corpus},
  author={Balke, Stefan and others},
  journal={Trans. Int. Soc. Music Information Retrieval},
  volume={9},
  number={1},
  year={2026}
}

@inproceedings{korzeniowski2025simple,
  title={Simple and Effective Semantic Song Segmentation},
  author={Korzeniowski, Filip and Vogl, Richard},
  booktitle={Int. Soc. Music Information Retrieval Conf. (ISMIR)},
  year={2025}
}

\documentclass[smc2026template.tex]{subfiles}

\ifSubfilesClassLoaded{%
  \usepackage{xr-hyper}%
  \externaldocument{smc2026template}%
}{}

\begin{document}

    \appendix
\onecolumn

\section{Scores according to the downstream segmentation algorithm}

\begin{figure*}[ht!]
    \centering
    \includegraphics[width=\linewidth]{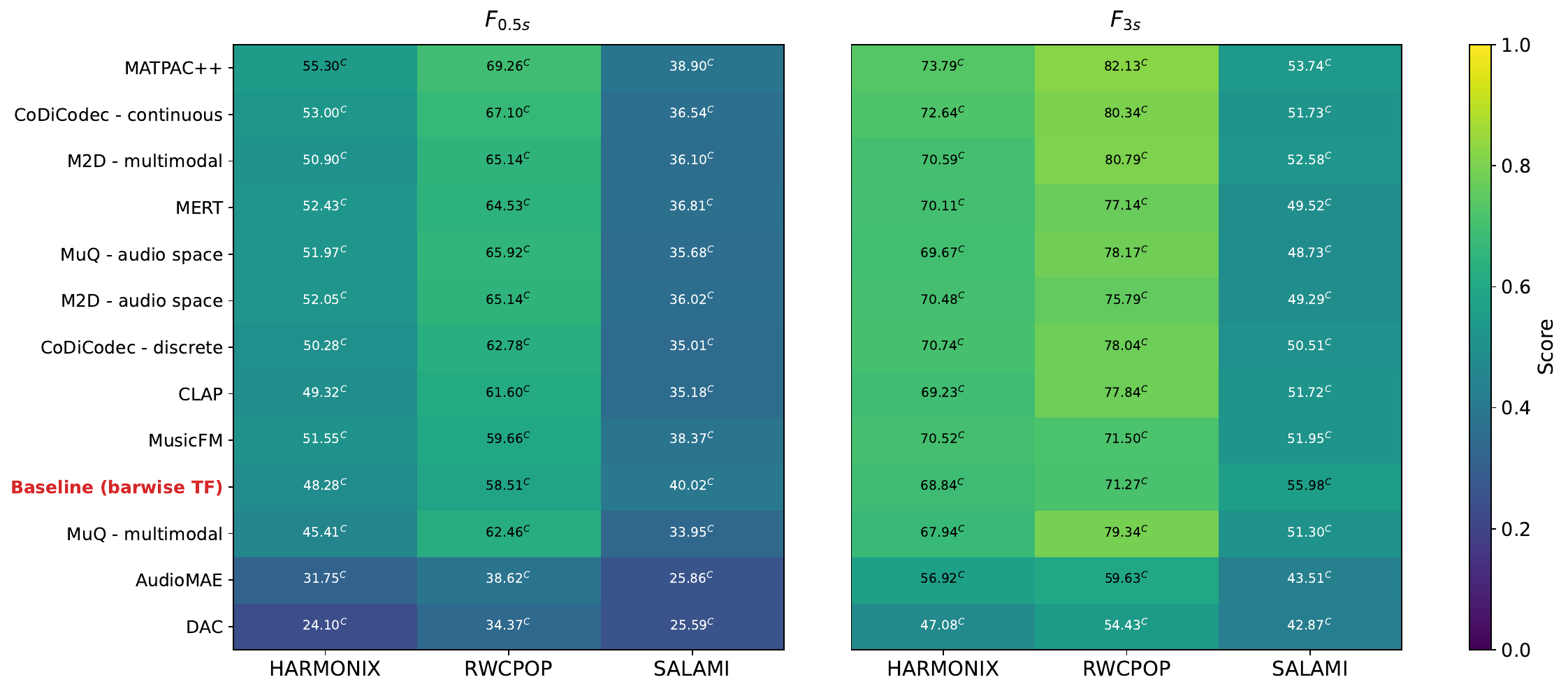}
    \caption{Best results obtained with all deep learning models, using the CBM segmentation algorithm. Rows are ordered by decreasing average of $\Fzf$ and $\Fth$. Hyperparameters are selected in the cross-dataset condition.}
    \label{fig:cbm_best}
\end{figure*}

\begin{figure*}[ht!]
    \centering
    \includegraphics[width=\linewidth]{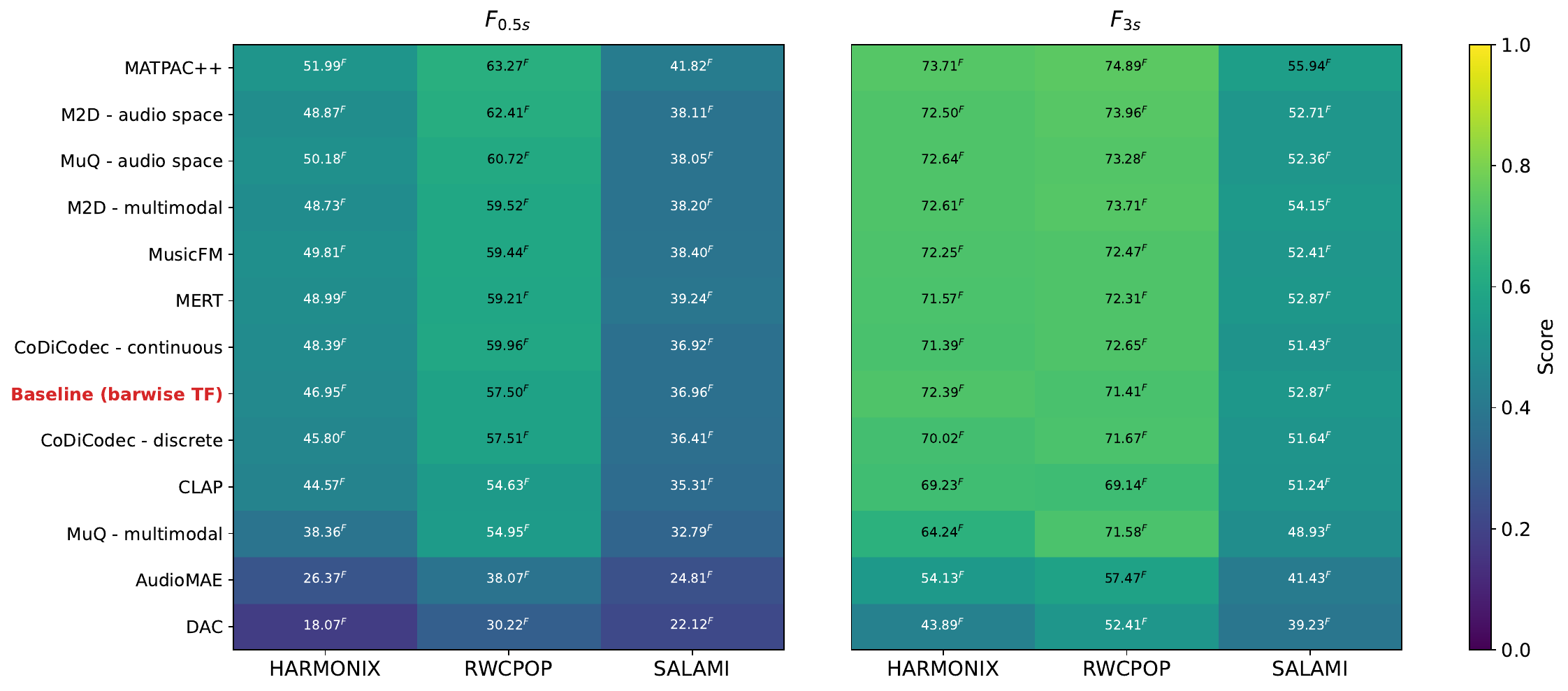}
    \caption{Best results obtained with all deep learning models, using the Foote segmentation algorithm. Rows are ordered by decreasing average of $\Fzf$ and $\Fth$. Hyperparameters are selected in the cross-dataset condition.}
    \label{fig:foote_best}
\end{figure*}

\begin{figure*}[ht!]
    \centering
    \includegraphics[width=\linewidth]{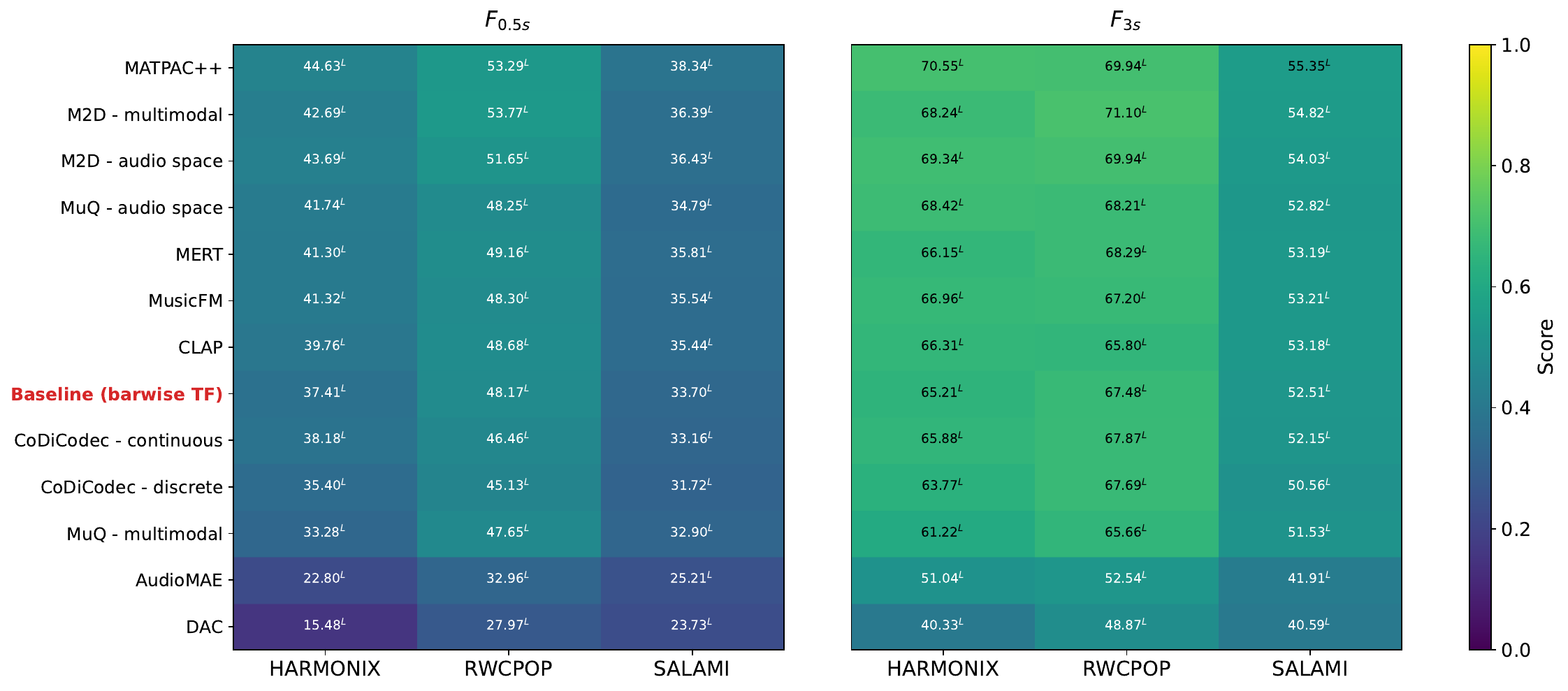}
    \caption{Best results obtained with all deep learning models, using the LSD segmentation algorithm. Rows are ordered by decreasing average of $\Fzf$ and $\Fth$. Hyperparameters are selected in the cross-dataset condition.}
    \label{fig:lsd_best}
\end{figure*}

In this section, we present three figures: one for each downstream segmentation algorithm (Figure~\ref{fig:cbm_best} for CBM, Figure~\ref{fig:foote_best} for Foote, and Figure~\ref{fig:lsd_best} for LSD). Hyperparameters are selected in the cross-dataset condition: as the best performing set per model, and across all datasets (as in Table 1. of the main article; see Sec.~\ref{sec:hyperparam_selection} of the main paper for more details). 

These results demonstrate consistent improvements when using deep models, with MATPAC++ and M2D achieving notably high scores. In contrast, DAC and AudioMAE are the worst-performing models. These findings confirm that using deep embeddings instead of standard features can substantially enhance segmentation performance, regardless of the downstream algorithm used.

\newpage
\section{Robustness of deep learning models across downstream segmentation algorithms}

\begin{figure*}[htb!]
     \centering
     \begin{subfigure}[b]{\textwidth}
         \centering
         \includegraphics[width=\textwidth]{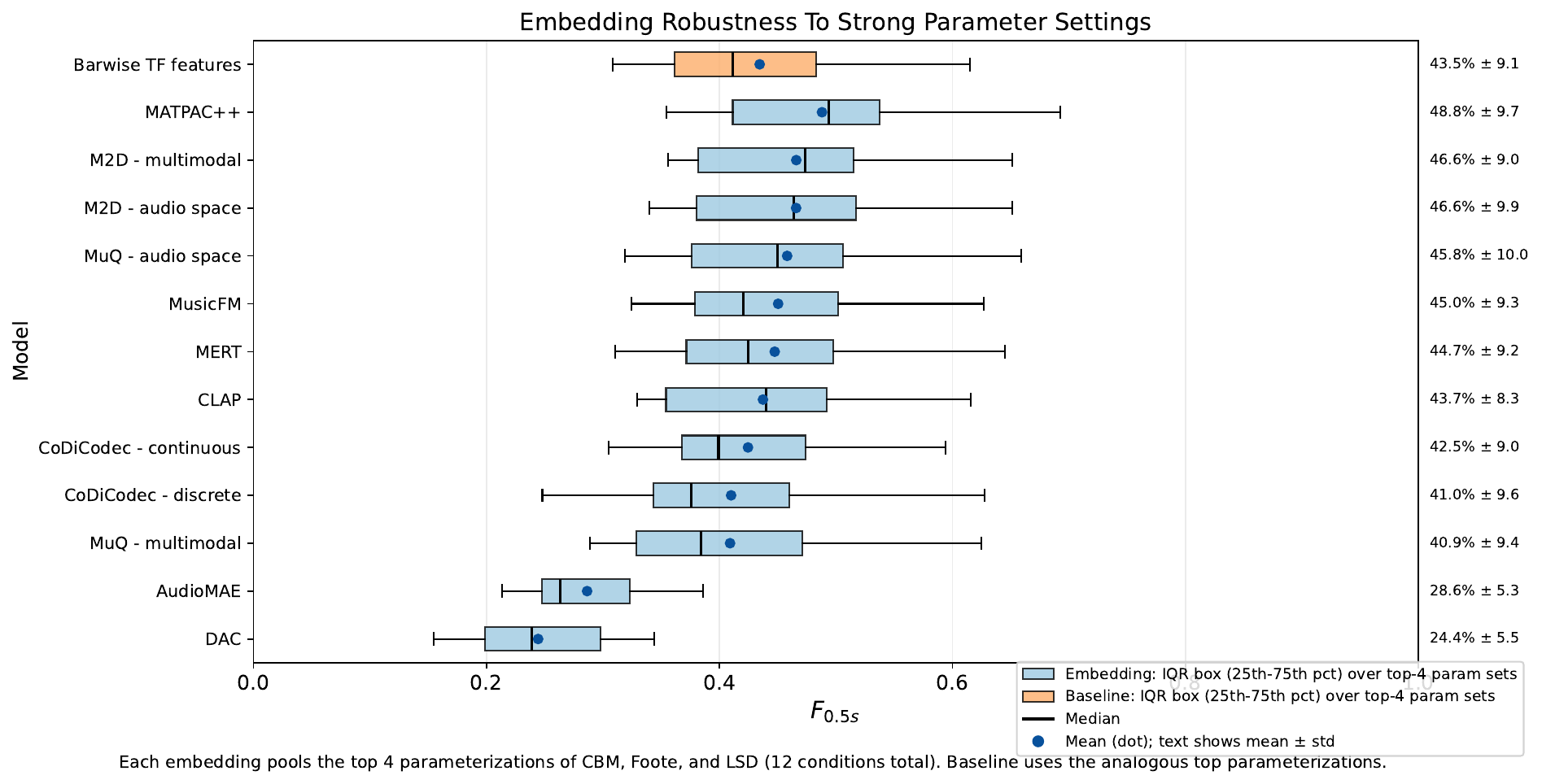}
         \caption{$\Fzf$}
     \end{subfigure}
     \hfill
    
     \centering
     \begin{subfigure}[b]{\textwidth}
         \centering
         \includegraphics[width=\textwidth]{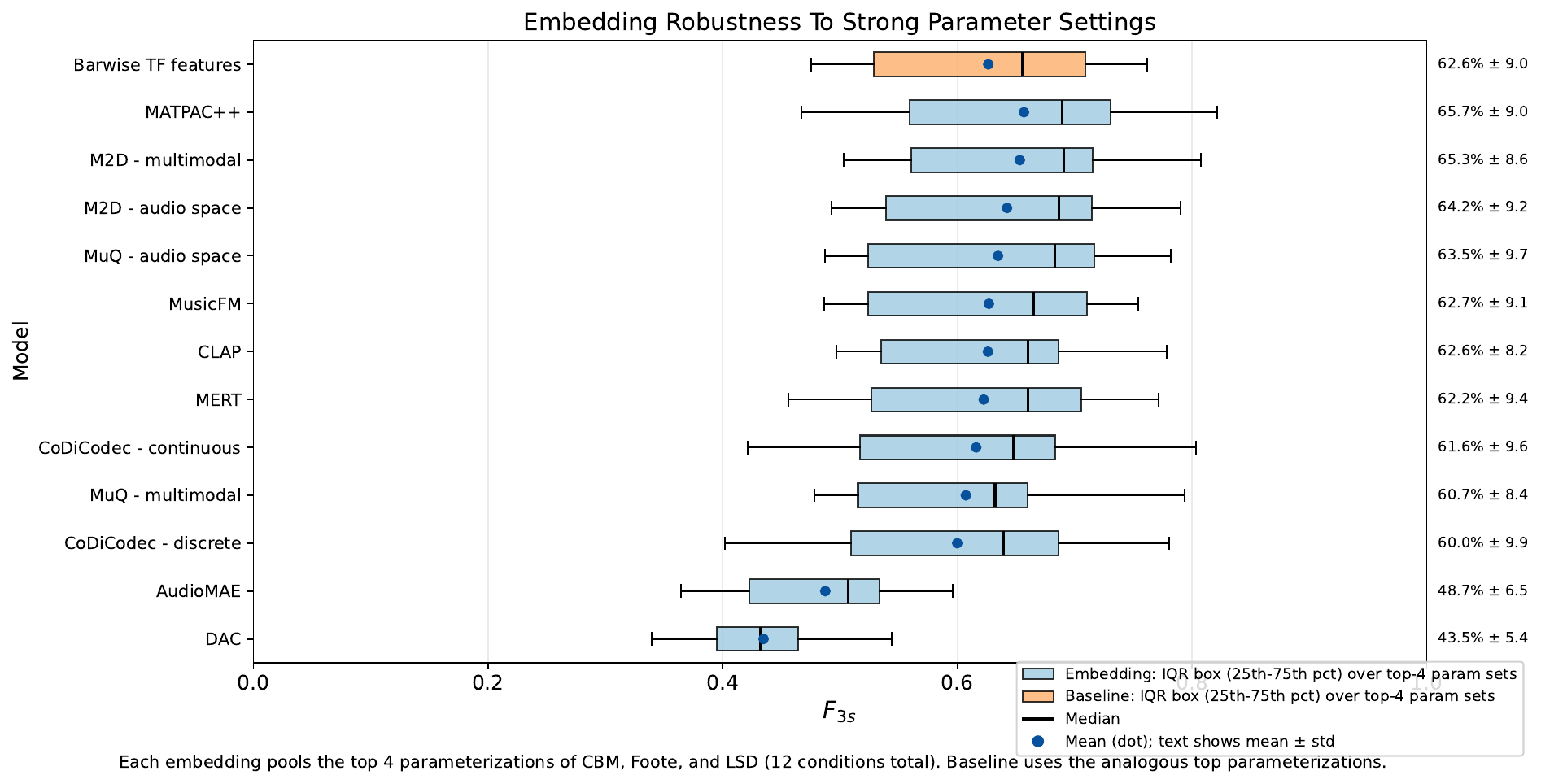}
         \caption{$\Fth$}
     \end{subfigure}
     \hfill

     \caption{Distribution of segmentation scores across downstream segmentation algorithms. Results are restricted to the four best-performing parameter configurations (averaged across all datasets, per model) for each algorithm, resulting in a total of 12 evaluated conditions per embedding.}
     \label{fig:robustness_embeddings}
\end{figure*}

This section evaluates the robustness of the embeddings across the various downstream segmentation algorithms. Figure~\ref{fig:robustness_embeddings} illustrates the distribution of segmentation scores for all embeddings across these algorithms. Rather than comparing all possible parameter configurations within our experimental setup, we selected the four best-performing configurations for each of the three studied algorithms, averaged across all datasets (leading to 12 configurations in total).

These results indicate a high sensitivity to the parameterization of downstream algorithms. However, results remain consistent with our previous findings: certain models consistently outperform others. Furthermore, the standard deviation of the baseline Barwise TF features is of the same order of magnitude as that of the deep models. This indicates that the observed performance variations stem primarily from the segmentation algorithms themselves, rather than from the embeddings.

\newpage
\section{Robustness of downstream segmentation algorithms to hyperparameter configurations}

\begin{figure*}[htb!]
     \centering
     \begin{subfigure}[b]{\textwidth}
         \centering
         \includegraphics[width=\textwidth]{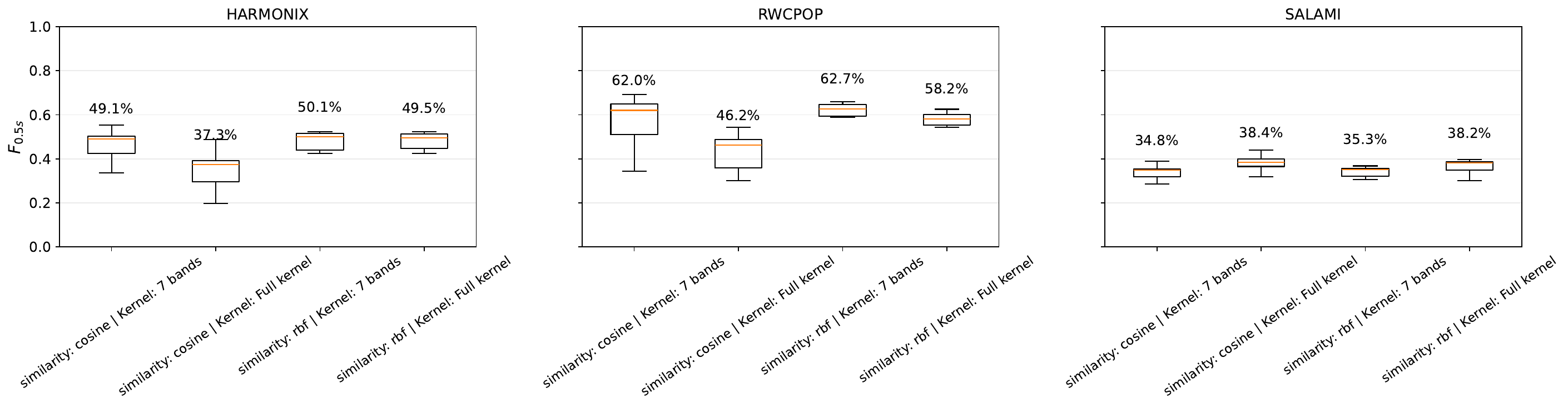}
         \caption{$\Fzf$}
     \end{subfigure}
     \hfill
     \centering
     \begin{subfigure}[b]{\textwidth}
         \centering
         \includegraphics[width=\textwidth]{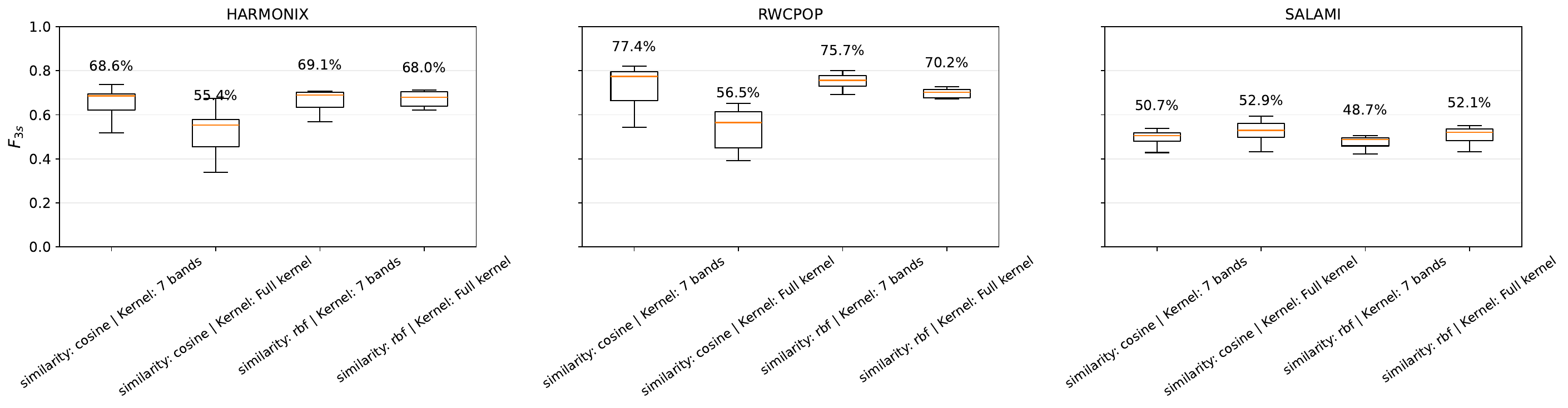}
         \caption{$\Fth$}
     \end{subfigure}
     \hfill
     \caption{Distribution of segmentation scores across all embeddings for the CBM algorithm, according to various parameter configurations. For each metric, each subfigure represents one dataset, respectively: Harmonix, RWC-Pop, and SALAMI.}
     \label{fig:robustness_cbm}
\end{figure*}

\begin{figure*}[htb!]
     \centering
     \begin{subfigure}[b]{\textwidth}
         \centering
         \includegraphics[width=\textwidth]{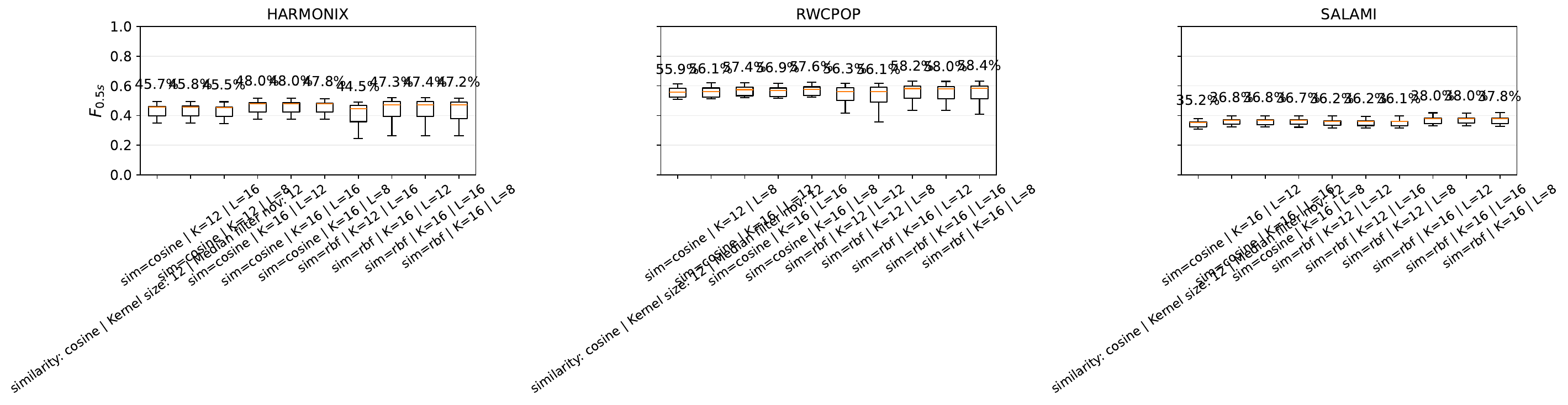}
         \caption{$\Fzf$}
     \end{subfigure}
     \hfill
     \centering
     \begin{subfigure}[b]{\textwidth}
         \centering
         \includegraphics[width=\textwidth]{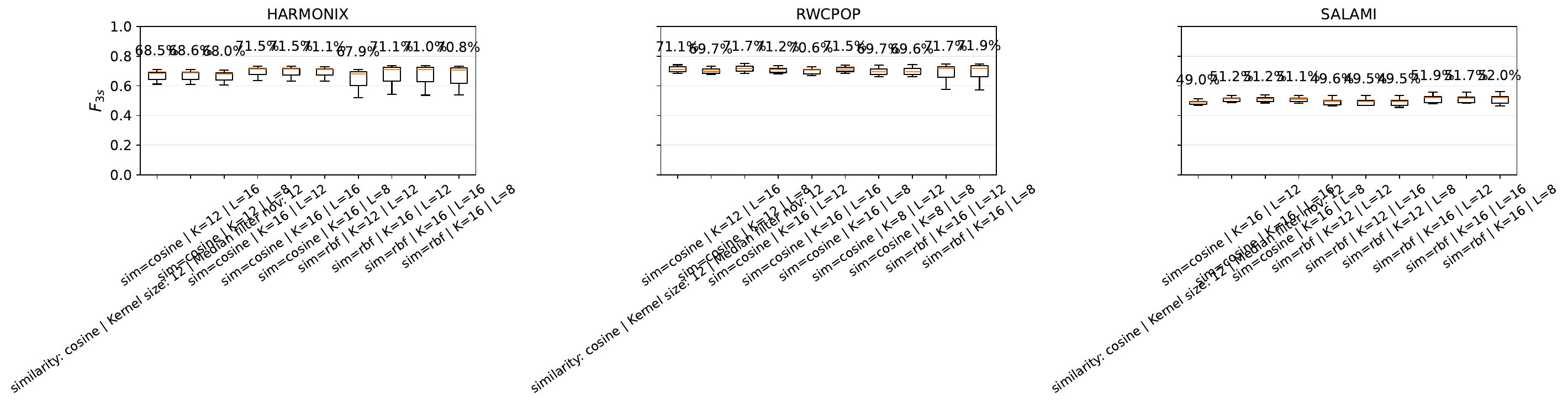}
         \caption{$\Fth$}
     \end{subfigure}
     \hfill
     \caption{Distribution of segmentation scores across all embeddings for the Foote algorithm, according to various parameter configurations. For each metric, each subfigure represents one dataset, respectively: Harmonix, RWC-Pop, and SALAMI.}
     \label{fig:robustness_foote}
\end{figure*}

\begin{figure*}[htb!]
     \centering
     \begin{subfigure}[b]{\textwidth}
         \centering
         \includegraphics[width=\textwidth]{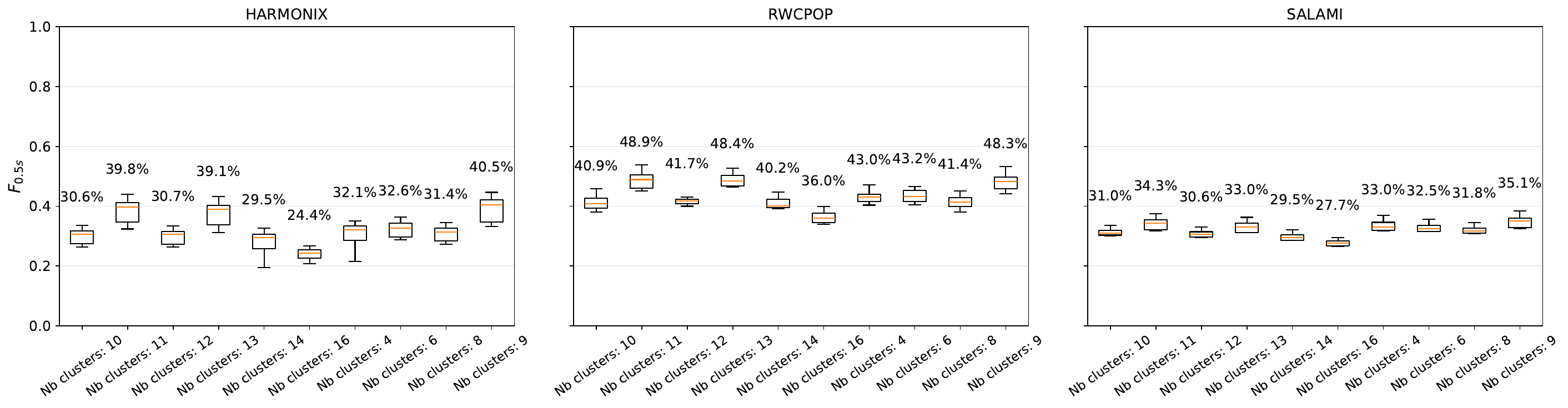}
         \caption{$\Fzf$}
     \end{subfigure}
     \hfill
     \centering
     \begin{subfigure}[b]{\textwidth}
         \centering
         \includegraphics[width=\textwidth]{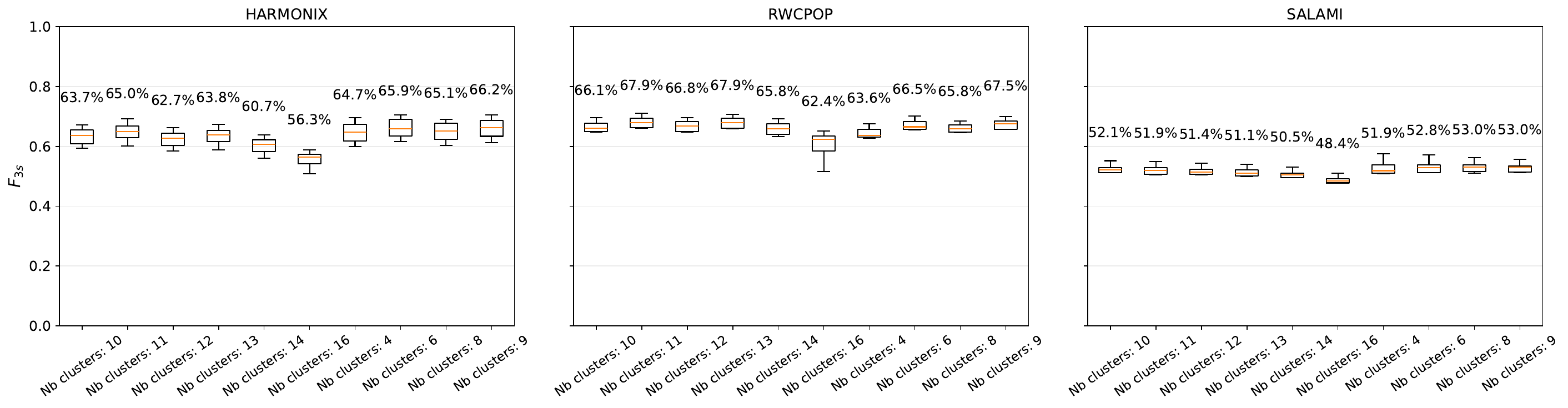}
         \caption{$\Fth$}
     \end{subfigure}
     \hfill
     \caption{Distribution of segmentation scores across all embeddings for the LSD algorithm, according to various parameter configurations. For each metric, each subfigure represents one dataset, respectively: Harmonix, RWC-Pop, and SALAMI.}
     \label{fig:robustness_lsd}
\end{figure*}

This section evaluates the robustness of the downstream segmentation algorithms across their specific parameter configurations. Figures~\ref{fig:robustness_cbm}, \ref{fig:robustness_foote}, and \ref{fig:robustness_lsd} present the distribution of scores for the CBM, Foote, and LSD algorithms, respectively, according to the different embeddings. Scores are presented per dataset.

Results demonstrate that the CBM algorithm is relatively stable when using RBF similarity, but yields lower and more erratic performance with cosine similarity. The Foote algorithm appears stable across the evaluated parameter configurations. Conversely, the LSD algorithm is highly dependent on the number of clusters, particularly concerning the $\Fzf$ metric. We did not investigate whether this instability arises from the LSD algorithm itself or from underlying assumptions that are incompatible with the barwise scale, which would make the algorithm ill-suited for this application. Confirming either explanation would require further experiments.

\end{document}

\end{document}